\documentclass[aps,prb,superscriptaddress,
 sd,%
 amsmath,amssymb,
 reprint,%
]{revtex4-1}

\usepackage[version=3]{mhchem} 
\usepackage{siunitx}[=2021-04-09] 
\usepackage{spin-textures}
\usepackage{graphicx}
\usepackage{dcolumn}
\usepackage{tabularx}
\usepackage{physics}
\usepackage{float}
\usepackage{amsmath,amssymb,amsthm,bm}
\usepackage{subfigure}

\begin{document}

\title{Domain-wall circulation gating and pinning via compositional engineering of cylindrical nanowires}
\author{Laura \'Alvaro-G\'omez}
\email{laura.alvaro@ucm.es}
 \affiliation{Dpto. de F\'isica de Materiales, Universidad Complutense de Madrid, 28040 Madrid, Spain.}
 
\author{Claudia Fern\'andez-Gonz\'alez}
\affiliation{Alba Synchrotron Light Facility, CELLS, 08290 Cerdanyola del Vall\`es, Barcelona, Spain.}

\author{Sandra Ruiz-G\'omez}
\affiliation{Alba Synchrotron Light Facility, CELLS, 08290 Cerdanyola del Vall\`es, Barcelona, Spain.}

\author{Nicolas Mille}
\affiliation{Synchrotron SOLEIL, l'Orme des Merisiers, Saint-Aubin, FR-91192 Gif-sur-Yvette Cedex, France.}

\author{Jean Christophe Toussaint}
\affiliation{Univ. Grenoble Alpes, CNRS, Institut N\'eel, 38000 Grenoble, France.}

\author{Rachid Belkhou}
\affiliation{Synchrotron SOLEIL, l'Orme des Merisiers, Saint-Aubin, FR-91192 Gif-sur-Yvette Cedex, France.}

\author{Michael Foerster}
\author{Eva Pereiro}
\author{Luc\'ia Aballe}
\author{N\'uria Bagu\'es}
\affiliation{Alba Synchrotron Light Facility, CELLS, 08290 Cerdanyola del Vall\`es, Barcelona, Spain.}

\author{Christophe Thirion}
\affiliation{Univ. Grenoble Alpes, CNRS, Institut N\'eel, 38000 Grenoble, France.}

\author{Daria Gusakova}
\affiliation{Univ. Grenoble Alpes, CNRS, CEA,  SPINTEC, 38000 Grenoble, France.}

\author{Aurelien Masseboeuf}
\affiliation{Univ. Grenoble Alpes, CNRS, CEA, SPINTEC, 38000 Grenoble, France.}

\author{Olivier Fruchart}
\affiliation{Univ. Grenoble Alpes, CNRS, CEA,  SPINTEC, 38000 Grenoble, France.}

\author{Lucas P\'erez}
 \email{lucas.perez@ucm.es}
 \affiliation{Dpto. de F\'isica de Materiales, Universidad Complutense de Madrid, 28040 Madrid, Spain.}
 \affiliation{IMDEA Nanociencia, Campus de Cantoblanco, 28049 Madrid, Spain.}

\begin{abstract}

Spintronic devices based on cylindrical magnetic nanowires have been proposed as a route toward three‑dimensional information storage and processing by harnessing the controlled motion of domain walls (DWs). In this geometry, the Bloch-point wall (BPW) is the characteristic DW type. It features a curling of magnetization around a Bloch Point singularity, a topology that has been predicted to enable propagation at high speed without undergoing a Walker breakdown. We demonstrate that compositional modulations of larger magnetization, which host spontaneous curling states while the rest of the wire remains axial, fulfil a dual role: they act as deterministic pinning sites and as active gates for BPW circulation. Combining magnetic transmission X‑ray microscopy and micromagnetic simulations, we reveal that DW–modulation interactions are the result of a superposition of long‑range magnetostatic repulsion and short‑range, circulation‑dependent exchange contributions: parallel circulation produces an attractive well, while opposite circulation generates a repulsive barrier.  This results in DW pinning at or near a compositional modulation. A DW’s internal circulation can be inverted upon traversing a modulation, enabling on‑demand switching of its circulation state. This dual mechanism demonstrates that both DW position and circulation can be tuned in a controlled manner, with potential implications for multistate spintronic devices.
\end{abstract}

\maketitle
\renewcommand{\tablename}{Table}


\section{Introduction}

Spin-based electronics has enabled commercial spintronic devices such as magnetic sensors in hard-disk drive read heads and magnetoresistive random-access memory (MRAM)  \cite{thomas2014perpendicular, bhatti2017spintronics, dieny2020opportunities}, demonstrating the power of electron-spin manipulation for ultrafast, nonvolatile operation. Spin-torque nano-oscillators (STNO) are also highly promising for radio-frequency applications, non-Boolean logic, and neuromorphic computing 
\cite{locatelli2014spin, awad2017long}. Over the past decades, planar spintronic platforms have utilized spin-transfer \cite{bib-PAR2008, ryu2013chiral} and spin–orbit torques\cite{miron2011perpendicular, emori2013current} to achieve reliable domain wall (DW) nucleation, propagation, and control, enabling proof-of-concept memory and logic devices, like the racetrack-memory concept \cite{thomas2011racetrack}, originally proposed in a three-dimensional (vertical) architecture to approach NAND-flash-like densities. Driven by this proposal and the maturity already achieved in 2D patterned magnetic structures, the field has been advancing towards three-dimensional architectures \cite{FernandezPacheco2017}, primarily motivated by their potential to achieve very large storage densities, making them an attractive, scalable alternative to existing memory technologies.

Cylindrical magnetic nanowires offer a versatile three-dimensional spintronic platform. Their strong uniaxial shape anisotropy stabilizes axial magnetization states \cite{bib-FRU2018d}. Moreover, magnetostatic interactions are much stronger in three-dimensional than in planar objects, which favors the formation of flux-closure magnetization distributions, commonly referred to as curling or vortex states \cite{Franchin2011, bib-FRU2019b, bib-FER2020,bib-FRU2021, bib-AG2022, FernandezRoldan2022,Moreno2022,Bran2023}. Curling textures may arise spontaneously and can be mobile, as in DWs with distinct internal structures: transverse‑vortex walls (TVWs), characterized by a transverse magnetization component with respect to the wire axis \cite{bib-FRU2015b}, and Bloch point walls (BPWs) \cite{bib-HER2002a,bib-FOR2002b,bib-FRU2014}, defined by azimuthal curling around a central magnetic singularity known as the Bloch point \cite{bib-DOE1968, bib-FEL1965, bib-MAL1979}.  Curling states can also be induced and stabilized at engineered discontinuities, such as wire terminations, diameter modulations \cite{bib-IGL2015,bib-BRA2016, bib-CHA2012,bib-SAL2013,bib-NAS2019, bib-ALL2009}, or chemical modulations (e.g.,   Fe\(_{20}\)Ni\(_{80}\)/Fe\(_{80}\)Ni\(_{20}\) \cite{Ruiz2020, bib-AG2022, FdezGonzalez2025-nanoletter}, Ni/Co \cite{bib-IVA2016b, bib-BER2017,Andersen2021}, or  Co/Cu\cite{bib-REY2016} interfaces),  where localized magnetic discontinuities favor flux-closure states with the form of curling magnetization. Axial charge currents generate Oersted fields whose curling symmetry couples directly to azimuthal magnetic textures. This enables BPW stabilization and ultra-fast motion  \cite{bib-FRU2019b}, eases DW nucleation \cite{bib-OTA2015} and controlled circulation reversal of BPWs \cite{bib-FRU2019b,bib-FRU2021,AlvaroGomez2025topo} and of engineered-modulations \cite{bib-AG2022}.

\begin{figure*}
\centering\includegraphics[width=0.7\textwidth]{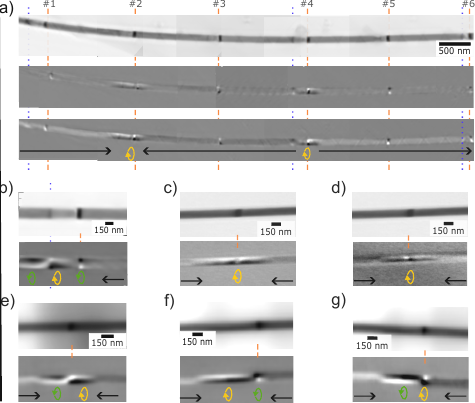}
\caption{Chemical modulations as domain wall pinning sites.  
(\textbf{a}) Composite transmission amplitude ptychographic reconstruction at the Fe L$_3$ edge of a 130 nm-diameter nanowire with 60 nm-long Fe$_{80}$Ni$_{20}$ chemical modulations. Top: transmission amplitude reconstruction. Middle: reconstructed  XMCD amplitude image measured at 0° tilt. Bottom: XMCD image with the sample tilted by 10°.  
(\textbf{b}, \textbf{e}) Transmission amplitude ptychographic reconstructions (top panels) and corresponding XMCD images (bottom panels) of a TVW located next to, or at, a chemical modulation, respectively.  
(\textbf{c}, \textbf{d}) TXM images at the Fe L$_3$ edge and Fe L$_2$ edge, respectively, together with the corresponding XMCD images (bottom panels), showing a BPW located at a modulation.  
(\textbf{f}, \textbf{g}) Transmission amplitude ptychographic reconstructions (top panels) and corresponding XMCD images (bottom panels) of a BPW located next to a chemical modulation for the two opposite senses of circulation, in both cases opposite one from another in the DW compared to the modulation.  
In all images, orange dashed lines mark the positions of the chemical modulations, and blue dotted lines indicate Fe-rich minor modulations. Nanowire diameter $d$ and modulation length $\ell$ are: (\textbf{a}, \textbf{b}) $d=\SI{130}{nm}$, $\ell=\SI{60}{nm}$; (\textbf{c}, \textbf{d}) $d=\SI{105}{nm}$, $\ell=\SI{100}{nm}$; (\textbf{e}, \textbf{f}, \textbf{g}) $d=\SI{120}{nm}$, $\ell=\SI{30}{nm}$.The circulation of magnetization in domain walls and modulations is indicated with yellow and green arrows.}
\label{fig:DW-pinning-experimental}
\end{figure*}

 Here we experimentally demonstrate that engineered compositional modulations —   nanoscale Fe-rich  zones (Fe\(_{80}\)Ni\(_{20}\)) alternated in micrometer-scale segments of Permalloy (Fe\(_{20}\)Ni\(_{80}\)) — act as DW traps and DW circulation gates that can invert the circulation of DWs when the DWs go through the modulation. We combine high-resolution X-ray microscopy (XMCD–TXM and ptychography) with micromagnetic simulations at rest and under the stimulus of a quasistatic applied axial magnetic field.  By clarifying the respective roles of dipolar and exchange interactions, we elucidate the distinct mechanisms governing domain wall pinning and propagation through compositional modulations. This dual mechanism is useful for information encoding not only through DW position but also via its circulation sense. The combination of both degrees of freedom provides access to multi-level states in cylindrical nanowires, which may be useful for 3D spintronic circuits.

\section{Experimental evidence }

\subsection{Domain wall pinning}
\label{sec:DW-deppining-motion}

In this section, we report on the interaction between domain walls (DWs) and periodic Fe-rich compositional modulations embedded in Permalloy nanowires, resulting in pinning effects. DWs were previously nucleated either by applying an external magnetic field of approximately \SI{1}{\tesla} perpendicular to the wire axis or by injecting a  \SI{10}{ns}-long current pulse with an amplitude of \SI{1e12}{\ampere\per\meter\squared}.  We 
image the magnetization states in the wires with either full field or scanning transmission X-ray microscopy, as well as with X-ray ptychography, all combined with X-ray Magnetic Circular Dichroism (XMCD). The combination of high spatial resolution and direct probing of magnetization provides unambiguous information about the DW type and inner structure.

\begin{figure}
\centering\includegraphics[width=1\columnwidth]{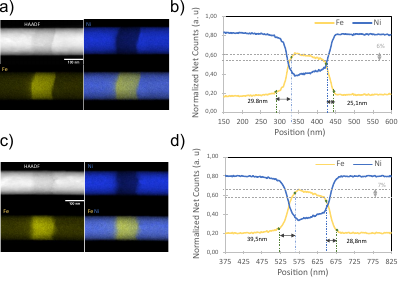}
\caption{STEM imaging and EDS elemental maps from two different modulations: modulation 1, a) and b); and modulation 2, b) and d). From both of them, HAADF-STEM image; individual Fe (yellow) and Ni (blue) elemental net count maps; combined Ni + Fe net count maps; and line profiles of the normalized net counts for Ni and Fe, integrated across the NWs width (130~nm for M1 and 108~nm for M2, respectively) are shown.}
\label{fig:TEM}
\end{figure}

\figref{fig:DW-pinning-experimental}{a} shows a transmission ptychographic reconstruction image of the amplitude at the Fe L$_3$ edge of a \SI{130}{nm}-diameter Permalloy nanowire containing periodic \SI{60}{nm}-long Fe\(_{80}\)Ni\(_{20}\) chemical modulations.  The Fe-rich modulations, indicated by orange dashed lines, appear as darker regions due to their higher X-ray absorption. A closer inspection reveals localized Fe depletion zones to the left of certain modulations, extending over several hundred nanometers and terminating in short, Fe-rich secondary modulations  (see blue dotted line, \eg, modulation $\#4$). These compositional inhomogeneities arise from instabilities during the electroplating process, where the nanowires grow from right to left, and were later mitigated by optimizing the synthesis protocol.  Despite such early-stage imperfections, the magnetic imaging already reveals essential features of DW pinning. 

The sharpness and homogeneity of the chemical modulations were investigated in detail using high-resolution Scanning Transmission Electron Microscopy. \figref{fig:TEM} shows the HAADF-STEM images together with the elemental maps and the corresponding composition profiles across the chemical modulations. As can be seen from the profiles, fitting the interface using a Boltzmann sigmoid function (for more details, see Section S1 of the Supporting Information) yields relatively sharp interfaces, with widths of approximately 30 nm, depending on the modulation. In addition, the Ni/Fe composition exhibits a variation of less than 6\% along the length of each modulation.
As will be demonstrated in the following sections, these variations in the size and composition of the chemical modulations are negligible compared with the characteristic length scale of the observed magnetic textures, which extends over several hundreds of nanometres. Therefore, from the perspective of their interaction with the magnetic textures, the chemical modulations can be considered homogeneous.

\begin{figure*}
\centering\includegraphics[width=0.9\textwidth]{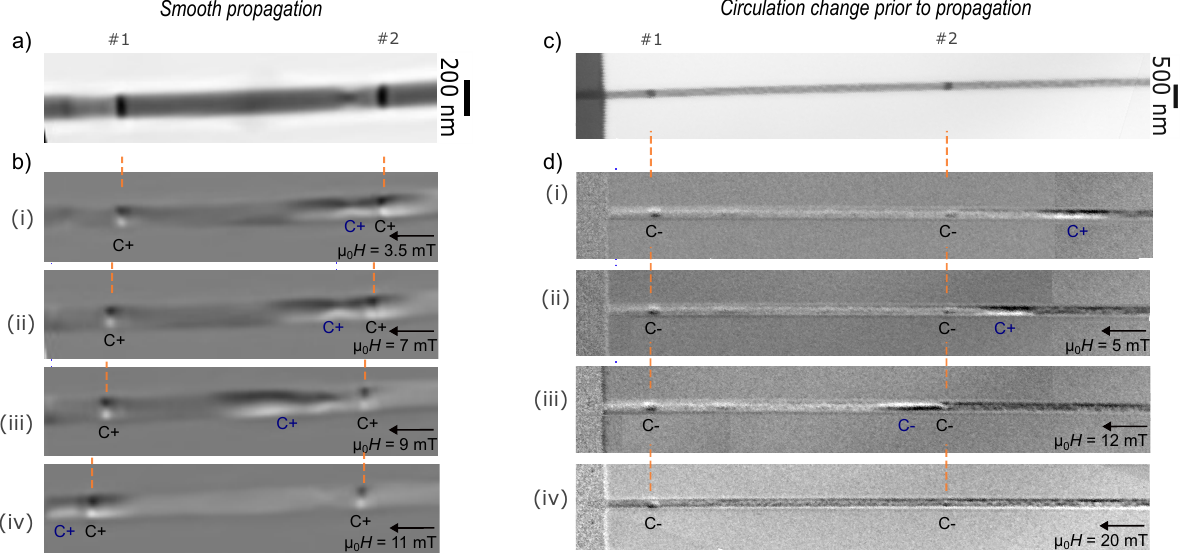}
\caption{Experimental evidence of chemical modulations acting as domain wall circulation gates for the situation of matching circulations (left panels) and antiparallel circulation (right panels). \textbf{a)} Transmission amplitude ptychographic reconstruction of a \SI{130}{nm} diameter nanowire with  \SI{60}{nm}-long Fe-rich chemical modulations.  \textbf{b)} Corresponding XMCD images acquired under a constant magnetic field (indicated in black), show BPW smooth propagation across chemical modulations that share the same circulation polarity ($C+ C+$).  \textbf{c)}  Full-field transmission amplitude image of a \SI{130}{nm} diameter nanowire with  \SI{100}{nm}-long Fe-rich chemical modulations.   \textbf{d)} Corresponding XMCD images taken under a constant magnetic field, indicated in black, show BPW polarity change (switching of its circulation) upon propagation across chemical modulation $\#2$ that exhibits opposite polarity ($C+ C-$).}

\label{fig:BPW-trough-modulation}
\end{figure*}
 
XMCD images are shown for two different geometries in \subfigref{fig:DW-pinning-experimental}{a}: the middle panel corresponds to the wire axis at \SI{0}{\degree} (perpendicular to the X-ray beam), and the bottom panel to a \SI{10}{\degree} tilt towards the beam. In the \SI{0}{\degree} configuration, the neutral contrast in the Permalloy segments confirms uniform axial magnetization, while the strong dark/bright contrast around the modulations indicates curling of magnetization around the axis \cite{bib-AG2022}. The tilted configuration also provides sensitivity to the axial component of the magnetization and reveals three magnetic domains (black arrows), separated by domain walls located close to modulations $\#2$ and $\#4$. At these positions, strong extended transverse black/white contrast is observed, which is characteristic of BPWs. This contrast can already be discerned in the \SI{0}{\degree} image, but the tilt allows their unambiguous identification by clearly separating domains with opposite axial magnetization. The walls are centered at the main modulations, while secondary modulations induce weaker curling, either parallel (modulation $\#2$) or antiparallel (modulation $\#4$) to that of the BPW circulation. Due to enhanced absorption in Fe-rich regions, XMCD contrast is stronger there but does not quantitatively reflect the component of magnetization parallel to the beam \cite{bib-FRU2015c}.

At the present wire diameters, DWs are much wider than the nanowire cross-section and always involve curling magnetization \cite{bib-FRU2015b}: Bloch point walls (BPWs) exhibit a uniform circulation around the axis, whereas transverse vortex walls (TVWs) display opposite curling on either side of their center. When a DW encounters a chemical modulation, its circulation may match that of the modulation, or it may be opposite. \figref{fig:DW-pinning-experimental} illustrates both cases for nanowires with different diameters and modulation lengths. When the DW and the modulation share the same circulation (\subfigref{fig:DW-pinning-experimental}{c–e}), BPWs and TVWs are  observed directly at the modulation, consistent with a potential-well interaction. In contrast, when the circulation is opposite, BPWs are often found adjacent to the modulation (\subfigref{fig:DW-pinning-experimental}{f–g}), suggesting repulsive interactions that prevent the wall from entering the modulation. For TVWs, the same trend is observed: if the circulation of the wall curling matches that of the modulation, the wall is stabilized at the defect (\subfigref{fig:DW-pinning-experimental}{e}), while opposite circulation suggests repulsion (\subfigref{fig:DW-pinning-experimental}{b}).

Overall, DWs are systematically located at or near chemical modulations rather than at random places along the wire. This suggests that the pinning/repulsion effects associated with modulations are stronger than those arising from  material inhomogeneities. We will further analyze these observations with the help of micromagnetic simulations in section~\ref{sec:discussion}.

\subsection{Domain wall circulation gate}
\label{sec:DW-deppining-motion} 
In the previous section we showed domain walls either residing at a modulation or remaining nearby, depending on whether their circulation matches or opposes that of the modulation. This raises the question of how a DW behaves when it is forced to propagate through a modulation, especially when its circulation is initially opposite to that of the defect.

To address this, we drove DW motion by applying an external axial magnetic field. \figref{fig:BPW-trough-modulation} summarizes two representative cases: to the left, a BPW propagates across two modulations with its circulation parallel to theirs ($C+C+$); to the right, the DW initially has opposite circulation to that of the modulation ($C+C-$).

In the $C+C+$ case, we examined a \SI{130}{nm}-diameter Permalloy nanowire containing two \SI{60}{nm}-long chemical modulations, labeled $\#1$ and $\#2$, separated by \SI{2}{\micro\meter} (\subfigref{fig:BPW-trough-modulation}{a}). Initially, a head-to-head BPW was located at modulation $\#2$, while no DW was present at modulation $\#1$ (\subfigref{fig:BPW-trough-modulation}{b.i}). When a DC axial magnetic field was applied along the wire axis, the DW left modulation $\#2$ at a depinning field of about \SI{9}{\milli\tesla} and moved towards the left (\subfigref{fig:BPW-trough-modulation}{b.iii}). At \SI{11}{\milli\tesla}, the DW has crossed the Permalloy segment and reached modulation $\#1$ (\subfigref{fig:BPW-trough-modulation}{b.iv}). These observations provide an experimental estimate of the depinning field for BPWs in the case of parallel circulation. Overall, these results show that, under circulation alignment, the modulations allowing the smooth BPW propagation once a critical field is exceeded.

In the $C+C-$ case, we examined a \SI{130}{nm}-diameter Permalloy nanowire with two \SI{100}{nm}-long chemical modulations, labeled $\#1$ and $\#2$, separated by \SI{6}{\micro\meter} (\subfigref{fig:BPW-trough-modulation}{c}). The initial XMCD image is shown in \subfigref{fig:BPW-trough-modulation}{d.i}. Both modulations display negative curling ($C^-$), while the BPW, located to the right of modulation $\#2$, exhibits positive circulation ($C^+$). When a \SI{5}{\milli\tesla} axial magnetic field was applied, the DW moved towards modulation $\#2$ but remained located just before it (\subfigref{fig:BPW-trough-modulation}{d.ii}). By gradually increasing the field (data not shown), the DW eventually crossed the modulation at \SI{12}{\milli\tesla}. The configuration imaged at this point shows that the DW has reversed its circulation to match that of the modulation ($C^-$) and is now situated just beyond it (\subfigref{fig:BPW-trough-modulation}{d.iii}). To verify that the observed circulation switching arose from the DW and not from a change in the modulation’s curling, the field was further increased to \SI{20}{\milli\tesla}, driving the DW towards the left contact. The resulting state (\subfigref{fig:BPW-trough-modulation}{d.iv}) confirms that both chemical modulations retained their original negative circulation ($C^-$). On this basis, modulations appear as circulation gates; i.e., they deterministically set the circulation of the wall to their own.

It should be noted that the XMCD-TXM images capture static magnetic configurations before and after field application. Although the analysis of multiple events strongly suggests that circulation reversal is governed by the chemical modulation, static snapshots cannot resolve sub-nanosecond dynamics during propagation. Directly tracking potential transient intermediate states would require future time-resolved or stroboscopic X-ray microscopy. To address this intrinsic experimental limitation in the particular case of domain wall transformation, we have performed a theoretical topological analysis at the end of the next section.

\section{Discussion: micromagnetic simulations}
\label{sec:discussion}

The experiments reported in the previous section demonstrate that the interaction of domain walls with chemical modulations affects both their position along the nanowire and the sign of their circulation. To clarify the mechanisms underlying these observations, we now turn to micromagnetic simulations. In particular, we evaluate the mutual interaction energy of a Bloch-point domain wall with a chemical modulation as a function of wire diameter, modulation length, and relative circulation. For this purpose, we simulate the motion of a DW under an applied magnetic field, gradually increasing the field strength in plateaus of \SI{2}{ns} duration and rising magnitude. 

\subsection{Energy landscape of a DW at a modulation}
\label{sec:simus-landscape}

Given the one-dimensional character of cylindrical nanowires, we reduce the domain wall energy landscape to a 1D profile, as in the classical Becker-Kondorski framework \cite{bib-KON1937}. In micromagnetic simulations, one common method to evaluate the energy landscape of a domain wall is to let it drift slowly in the absence of external forces, under high damping conditions. This ensures that the system remains in a quasi-static regime throughout the process. Such an approach was successfully applied to diameter-modulated nanowires in previous work \cite{bib-FRU2020}. However, spontaneous drift may be insufficient to map energy landscapes with non-monotonic profiles, especially for repulsive parts. In such cases, we apply an axial magnetic field to push the DW in a controlled manner while maintaining its internal structure nearly unchanged. In practice, to maintain quasi-static conditions and minimize dynamic distortions, the damping constant is set to $\alpha = 1$. 

\begin{figure}[h]
\centering\includegraphics{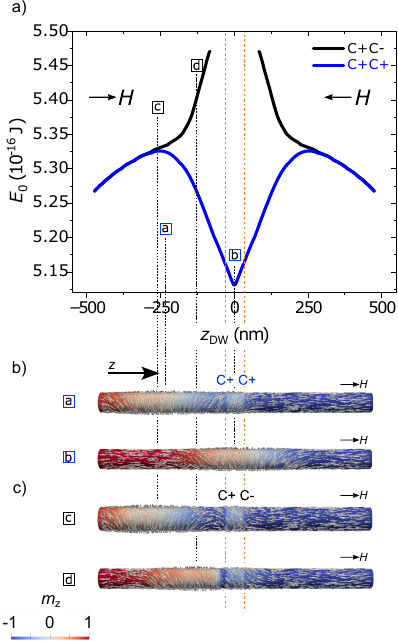}
\caption{Energy landscape due to DW-modulation interaction. \textbf{a)} Internal energy $E_0$ versus head-to-head BPW position (z$_{\mathrm{DW}}$) in a \SI{130}{nm} diameter Permalloy nanowire with a \SI{60}{nm}-long chemical modulation centered at $z=0$. Orange dashed lines indicate the two interfaces of the modulation. The wall is driven towards the modulation by an axial quasi-static magnetic field of rising magnitude oriented along $\unitvectz$ or $-\unitvectz$ for z$_{\mathrm{DW}} <0$ or z$_{\mathrm{DW}} >0$, respectively. The blue curve indicates parallel circulations $C+$$C+$  of the BPW and modulation, while the black curve represents antiparallel $C+$$C-$ circulations. \textbf{b)} and \textbf{c)} Outer view configurations obtained with micromagnetic simulations for selected wall positions (z$_{\mathrm{DW}}$) for parallel and antiparallel circulations, respectively. The color code represents $m_{\mathrm{z}}$.}
\label{fig:DW-pinning-landscape}
\end{figure}

We consider a \SI{130}{nm}-diameter Permalloy (Fe\(_{20}\)Ni\(_{80}\)) nanowire with a \SI{60}{nm}-long Fe\(_{80}\)Ni\(_{20}\) chemical modulation centered at \( z= 0 \).  The reduced diameter compared with experiments was chosen to lower computational cost while preserving the essential physics, as the key phenomena observed—domain wall pinning, repulsion, and circulation switching—remain robust across this intermediate diameter range.
We considered two initial situations: parallel and antiparallel curling circulations between the BPW and the modulation, labeled as \( C+C+ \) and \( C+C- \), respectively. \( z_{\mathrm{DW}} \) denotes the axial position of the domain wall along the nanowire. The initial position of the DW is   \( z_{\mathrm{DW,i}} = \pm \SI{500}{nm} \). The domain wall is pushed towards the modulation by applying an external magnetic field along \( \unitvectz \) or \( -\unitvectz \), depending on whether \( z_{\mathrm{DW,i}} < 0 \) or \( z_{\mathrm{DW,i}} > 0 \), respectively. The initial BPW circulation is chosen to be positive with respect the direction of motion, ensuring that it remains stable and does not switch spontaneously during the dynamics~\cite{bib-THI2006}.  

\subfigref{fig:DW-pinning-landscape}{a} displays the evolution of the internal energy as a function of \( z_{\mathrm{DW}} \), defined as the sum of dipolar and exchange contributions \( E_0 = \Ed + \Eex \). The value of \( z_{\mathrm{DW}} \) is determined as the intersection point of the three isosurfaces of the magnetization field: \( m_{\mathrm{x}} = 0 \), \( m_{\mathrm{y}} = 0 \), and \( m_{\mathrm{z}} = 0 \)~\cite{bib-AND2014b}.  The blue curve represents the case of parallel circulations \( C+C+ \), and the black curve represents the situation of antiparallel circulations \( C+C- \).  \subfigref{fig:DW-pinning-landscape}{b} and \subfigref{fig:DW-pinning-landscape}{c} show outer views of the micromagnetic simulation at selected positions. 
 
 At a long distance from the modulation, typically $\SI{250}{nm}$ or more, both circulation configurations yield overlapping energy profiles, revealing a repulsive interaction. Indeed, at long range, the interaction is dominated by magnetostatics and is insensitive to the relative circulation between the BPW and the modulation. This is consistent with the known fact that circulations of curling structures do not interact with one another via dipolar interactions\cite{bib-WYS2017}, as the magnetic potential does not depend on the sign of circulation. The repulsion results from the interaction of the positively-charged head-to-head BPW with the dipolar field of the modulation, whose positively-charged interface is the one closest to the wall\cite{bib-AG2022}. A similar repulsion would occur for a tail-to-tail BPW encountering a negatively-charged modulation interface.

 At shorter separations, however, the interaction depends on the relative circulations of wall and modulation: it becomes increasingly repulsive for opposite circulations, while it becomes attractive for identical circulations. This results from the direct exchange interaction between the two magnetization textures: for matching circulations, the longitudinal magnetization gradients decrease, minimizing exchange energy and producing a local potential well (blue curve). Conversely, opposite circulations increase longitudinal magnetization gradients, raising exchange energy and resulting in an energy barrier  (black curve). This behavior is consistent with previous reports on the interaction of curling end structures in short nanotubes \cite{bib-WYS2017}, and matches the experimental observations presented earlier, where DWs with the same circulation as the modulation were found at the center of the defect [Fig.~\ref{fig:DW-pinning-experimental}(c)], while those with opposite circulation stopped before crossing it [Fig.~\ref{fig:DW-pinning-experimental}(f,g)].

\subsection{Propagation of DWs through modulations}

\subsubsection{The case of identical circulations -- Smooth propagation}
\label{sec:BPW-propagation-field} 

We now examine how BPWs propagate across chemical modulations under an axial field when the wall and the modulation share the same circulation sense. The initial state is a head-to-head BPW located at $z_{\mathrm{DW}} =\SI{-500}{nm}$, which we push towards the modulation by applying a positive magnetic field.

\begin{figure*}
\centering\includegraphics{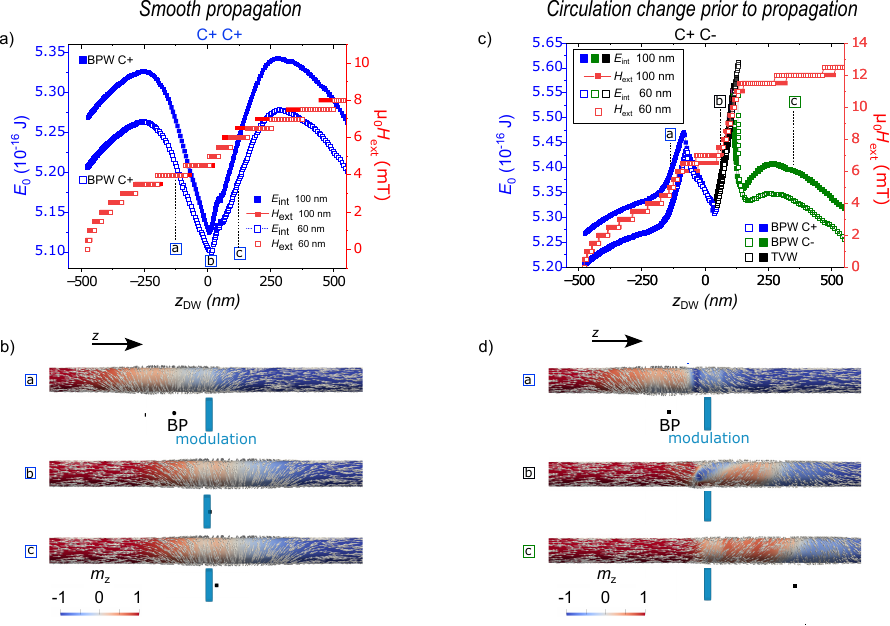}
\caption{Micromagnetic simulations of DW circulation gating by chemical modulations for matching (left panels) vs. antiparallel circulations (right panels).  \textbf{a)},  \textbf{c)}: Internal energy $E_0$ versus BPW position ($z_{\mathrm{DW}}$) evolution. As DWs change type while propagating, we use colors to identify their type over time. Blue, black, and green symbols represent the DW configurations of BPW \( C+ \), TVW, and BPW \( C-\), respectively. Open and full symbols correspond to a chemical modulation length of \SI{60}{nm} and \SI{100}{nm}, respectively. Red symbols show the magnitude of the axial magnetic field, for every location of the DW, applied in plateaus of duration \SI{2}{ns} and rising magnitude.  \textbf{b)},  \textbf{d)}: Outer  micromagnetic views at selected $z_{\mathrm{DW}}$, for a modulation length of  \SI{60}{nm}. The black square corresponds to the Bloch point position and the elongated light blue rectangle to the position of the chemical modulation.
 }
\label{fig:simulations-BPW-propagation-modulation}
\end{figure*}

\subfigref{fig:simulations-BPW-propagation-modulation}{a} shows the evolution of the internal energy  of the system as a function of the DW axial position  $z_{\mathrm{DW}}$, under a gradually increasing axial magnetic field. Two modulation lengths are compared: \SI{60}{nm} (open blue symbols) and \SI{100}{nm} (filled blue symbols), with red symbols indicating the field strength for each DW position, applied in \SI{2}{ns}-long plateaus.

Both energy profiles exhibit qualitatively similar features: a repulsive tail at long distances and a local energy minimum at the modulation site, consistent with the static energy landscapes shown in \subfigref{fig:DW-pinning-landscape}{a}. Initially, the DW is located in the repulsive region to the left of the modulation, and begins to move when the applied field allows it to overcome the axial energy gradient. It is worth noting that the wall already displays a continuous drift during each \SI{2}{ns} field plateau near the top of the barrier (around $z_{\mathrm{DW}}=\SI{-250}{nm}$). Thus the applied field value, of the order of \SI{3.5}{\milli\tesla}, corresponds to an upper estimate of the maximum repulsive field, largely independent of the modulation length.

Once the DW enters the attractive region, it accelerates before becoming pinned near the bottom of the energy well. It remains stationary over a broad field range, as indicated by the steep rise of the red curve. Eventually, it depins and resumes motion towards the right. The depinning fields are approximately \SI{5}{mT} and \SI{5.5}{mT} for the short and long modulations, respectively. The hierarchy of these two values reflects the maximum slope of the energy curves, in qualitative agreement with the Becker–Kondorsky model.

For the longer modulation, the entire internal energy profile is shifted upward, even far from the modulation, reflecting the higher energetic cost of maintaining curling over a longer Fe-rich segment~\cite{bib-AG2022}. This shift is partially compensated when the wall is pinned at the modulation, since most of the BPW curling fits inside the modulation and benefits from the exchange energy already required by the composition modulation and the depinning field is higher.

Finally, the energy maximum at $z_{\mathrm{DW}}> 0$ is slightly higher than at $z_{\mathrm{DW}}< 0$, in contrast with the symmetric energy landscape computed in \subfigref{fig:DW-pinning-landscape}{a}. This asymmetry likely stems from subtle changes in the internal structure of the BPW under the applied field, which differ upon entering versus exiting the modulation. 

As the peripheral curling and the Bloch point may follow different dynamics during the motion of a BPW \cite{bib-HER2004a}, it is interesting to track the Bloch point during the propagation/pinning/depinning process, which is illustrated in \subfigref{fig:simulations-BPW-propagation-modulation}{b}. An external view of the magnetization field at the wire surface is shown with white arrows representing the local magnetization. The background color codes the axial component of magnetization $m_{\mathrm{z}}$, and thus the location of the wall, as it results from the associated head-to-head charges \cite{bib-THI2006}. Below each view, the black square indicates the axial position of the Bloch point.  Both match which shows that a BPW propagates smoothly through chemical modulation when its circulation senses are parallel.\\

\subsubsection{\label{sec:BPW-chirality-polarizer}The case of opposite circulations -- Modulations as a polarizing gate}
\label{sec:BPW-chirality-polarizer}

In sharp contrast to the parallel-circulation scenario, the internal energy landscape of the antiparallel configuration 
displays a discontinuity and a pronounced barrier, as shown in \subfigref{fig:simulations-BPW-propagation-modulation}{c}. Simulations were carried out for modulation lengths of \SI{60}{nm} and \SI{100}{nm}, with the field amplitude increased stepwise in \SI{2}{ns} plateaus (red symbols).

In this case, the DW cannot propagate smoothly through the modulation. Instead, it undergoes a sharp jump in internal energy. It requires substantially higher magnetic fields to overcome the barrier: approximately \SI{11}{mT} for the shorter modulation and up to \SI{11.5}{mT} for the longer one.

Micromagnetic snapshots in \subfigref{fig:simulations-BPW-propagation-modulation}{d} illustrate the underlying mechanism. At the initial stage (a), the BPW and the modulation exhibit antiparallel circulations. As the wall approaches the modulation, it shrinks and its energy rises (blue symbols in \subfigref{fig:simulations-BPW-propagation-modulation}{c}). At point (b), the associated Bloch point vanishes, resulting in a transient TVW configuration (black symbols). At point (c), a Bloch point reappears to the right of the modulation, where the BPW and the modulation now share the same configuration, confirming that the wall has reversed its circulation. The resulting BPW with negative circulation (green symbols in \subfigref{fig:simulations-BPW-propagation-modulation}{c}) subsequently depins and drifts into the potential well defined by parallel circulations.

These results demonstrate that transmission across an antiparallel modulation does not proceed by simple drift but instead involves a topological transformation sequence (BPW $\rightarrow$ TVW $\rightarrow$ BPW of opposite circulation). Furthermore, the barrier height — and thus the required depinning field — increases with modulation length. 

An extended scaling study presented in the Supplementary Information demonstrates that the qualitative behavior described here—namely circulation-selective pinning and deterministic switching—remains robust across the experimentally relevant range of nanowire diameters and modulation lengths.

\begin{figure}
\centering\includegraphics{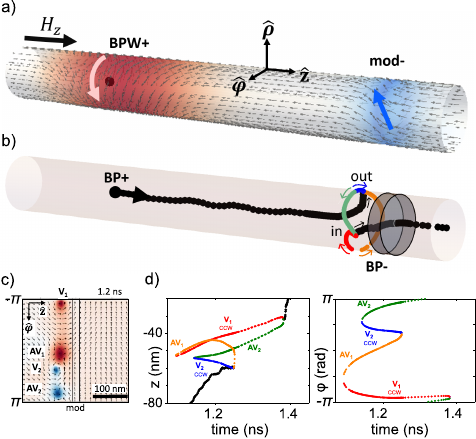}
\caption{Topology of field-driven BPW circulation switching prior to propagate through a chemical modulation of antiparallel circulation. \textbf{a)} Outer micromagnetic view of the initial state, with a BPW of positive circulation and a modulation of negative circulation with respect to $\unitvectz$. The color code represents $m_{\mathrm{\varphi}}$. (b) Black symbols: Bloch point trajectory while crossing the modulation under an axial field of \SI{50}{\milli\tesla}. Color: generated vortex/antivortex pairs on the surface. (c) Unrolled surface of the wire showing the vortex/antivortex pair formation to the left of the modulation. The color code represents $m_{\mathrm{\rho}}$. (d) Temporal evolution along $\unitvectz$ and $\unitvectvarphi$ of Bloch point (black) and vortex/antivortex pairs (colored curves). }
\label{fig:BPW-sw-under-field-topology}
\end{figure}

To better understand the microscopic mechanism underlying the domain wall transformation, we performed a topological analysis~\cite{AlvaroGomez2025topo}. To mimic realistic conditions, the simulations were carried out with a damping parameter $\alpha=0.02$ and under an axial magnetic field of \SI{50}{\milli\tesla}. \subfigref{fig:BPW-sw-under-field-topology}{a} shows the outer micromagnetic view of the initial state. The color code represents the azimuthal magnetization: red denotes the BPW curling along $+m_{\upvarphi}$, while blue corresponds to the chemical modulation magnetized along $-m_{\upvarphi}$. The tracking of volume and surface singularities was performed using a dedicated post-processing tool. The three-dimensional view in \subfigref{fig:BPW-sw-under-field-topology}{b} displays the trajectory of the Bloch point (black line) together with surface vortex–antivortex pairs (colored lines). The grey disk marks the chemical modulation. Far from the modulation, the Bloch point propagates steadily along the nanowire axis, as expected for DW motion in homogeneous nanowires~\cite{bib-HER2016}. As it approaches the modulation, two surface vortex–antivortex pairs with opposite polarities nucleate, split, and start traveling azimuthally on the wire surface. This process, associated with local regions of radial magnetization reaching unity, does not yet involve a change in topology.  

\subfigref{fig:BPW-sw-under-field-topology}{d} provides a detailed view of the subsequent transformation by plotting the longitudinal and azimuthal trajectories of all topological objects as a function of time. The first nucleated pair, V$_1^+$–AV$_1^+$, exhibits positive polarity (red and orange lines), while the second pair, V$_2^-$–AV$_2^-$, carries negative polarity (blue and green lines). In both cases, the vortex and antivortex separate azimuthally. The polarity of the first pair can be rationalized as it matches that of the head-to-head BPW and the sense favored by the chirality of the Landau–Lifshitz–Gilbert equation via $\vect{H}\times\vect{M}$. The appearance of the second pair with opposite polarity is likely driven by the reduction of global magnetostatic energy. The critical stage occurs when the Bloch point reaches the surface and annihilates with vortex V$_2^-$ and antivortex AV$_1^+$ (labeled $out$). At this point, only V$_1^+$ and AV$_2^-$ remain, separated far apart on the surface and the wall has the topology of a TVW. Their eventual annihilation leads to the re-injection of a Bloch point with negative polarity into the volume (labeled $in$), completing the topological transformation back to a BPW and allowing the DW to propagate across the modulation.

This mechanism closely mirrors the one reported by~\citealt{bib-FRU2021} for BPW switching in a homogeneous nanowire driven by an antiparallel Oersted field, as both share the same topological features. An analogy can be drawn with the classical distinction between coherent reversal and nucleation–propagation mechanisms in magnetization reversal: switching the BPW circulation via a homogeneous increase of $m_{\rho}$ to $\pm 1$ along the entire wire periphery would entail a prohibitive dipolar energy cost. Instead, we observe the local nucleation of a region with $m_{\rho} = \pm 1$, which then propagates azimuthally around the wire through the motion of surface vortex–antivortex pairs in opposite directions.  

When comparing BPW switching in homogeneous and chemically modulated systems, important differences emerge. While in the work by~\citealt{bib-FRU2021} this topological process is driven purely by inertia in Landau-Lifshitz-Gilbert micromagnetic dynamics, the driving force in the present case is exchange stiffness, giving rise to the transformation even for damping up to unity, for which inertia is largely inexistent. The unrolled map in \subfigref{fig:BPW-sw-under-field-topology}{c} reveals that vortex–antivortex pairs always nucleate on the same side of the BPW, specifically between the wall interface and the chemical modulation. This contrasts with the case of Oersted -field-induced switching in homogeneous nanowires, where one vortex or antivortex of the same polarity appears on each side of the wall~\cite{bib-FRU2021}. In the modulated system, the strong exchange interaction with the modulation of opposite circulation acts as an effective field but localized only on one side of the DW.

In summary, the chemical modulation not only functions as a circulation gate but also mimics an Oersted-field-like effect, enabling localized and asymmetric switching mediated by exchange interactions. This provides new opportunities for deterministic control of domain wall topology. One may wonder why it is the chemical modulation, a micromagnetic object only \SI{20}{nm} long, that forces the circulation of a BPW exceeding \SI{100}{nm} in length to switch, and not the reverse. We attribute this asymmetry to the stiffness of the curling state inside the modulation, imposed by the strong magnetization mismatch at its two interfaces. This robustly stabilizes the circulation within the modulation, closing the magnetic flux and minimizing magnetostatic energy, so that the BPW adapts to the modulation rather than the opposite. 

To assess the viability of this gating mechanism against ambient fluctuations, we examine the effective energy barriers (${\Delta E}\sim{10}^{-18}$~J) extracted from the micromagnetic landscapes (Figs. 3 and 4). Compared to the thermal energy at room temperature ($k_BT\approx4.14\times{10}^{-21}$~J), the normalized barrier is $\Delta E/k_B T \sim 240 $. Providing that, in the memory industry, a ratio of 40 to 60 is considered sufficient to prevent switching for several years, the Néel–Arrhenius analysis indicates that the system is highly resilient against thermally activated switching. Furthermore, the experimental operational window (approx. 10–30 mT) ensures that typical environmental field noise ($\ll 1$ mT) cannot trigger unintended transformations, confirming the structural robustness of the proposed topological gate.

\section{Conclusion}
\label{sec:conclusion}

In this work we investigated the interaction between Bloch-point domain walls and longitudinal chemical modulations in cylindrical magnetic nanowires, both characterized by a circulation of magnetization around the wire axis. Using a combination of magnetic imaging and micromagnetic simulations, we showed that the interaction depends critically on their relative position and circulation sense, giving rise to a non-monotonic energy landscape. At long range, beyond the domain-wall width, the interaction is purely magnetostatic—always repulsive and independent of circulation. At shorter distances, however, exchange dominates: identical circulations produce an attractive potential well that pins the DW at the modulation, while opposite circulations generate strong repulsion.  

When driven across a modulation by an external longitudinal field, the DW response depends on circulation matching. For identical circulations the wall depins smoothly, whereas for opposite circulations and sufficiently narrow modulations its circulation reverses. This reversal involves the annihilation and subsequent renucleation of the Bloch point, analogous to transformations previously reported in homogeneous nanowires. Overall, our results highlight the key role of relative circulation in the interaction of 3D magnetization textures, and establish high-magnetization chemical modulations as effective circulation gates for Bloch-point domain walls.

Finally, this study focuses solely on the fundamental physics of isolated nanostructures to uncover the intrinsic mechanisms of domain wall pinning without interwire interactions. While template-assisted electrodeposition enables low-cost synthesis of cylindrical nanowires with tailored chemical modulations, full device-level integration remains a major technological hurdle. Developing scalable integration protocols and demonstrating multi-bit functionality are beyond the scope of this study, aimed to establish domain wall polarity gating as a foundational building block for future three-dimensional spintronic architectures.

\section*{Methods}
Permalloy (Fe\(_{20}\)Ni\(_{80}\)) cylindrical nanowires with Fe-rich chemical modulations were synthesized by single-bath template-assisted electrochemical deposition. Nanoporous anodic aluminum oxide (AAO) templates were prepared by hard anodization of Al disks (Goodfellow, \SI{99.999}{\%} in purity) in a water-based solution of oxalic acid (\SI{0.3}{M}) and ethanol (\SI{0.9}{M}), applying \SI{140}{V} of anodization voltage at \SIrange{0}{1}{\celsius} for \SI{2.5}{h}. The remaining Al was etched with an aqueous solution of CuCl$_2$ (\SI{0.74}{M}) and HCl (\SI{3.25}{M}), the oxide barrier was removed and the pores opened to a final diameter of $\approx$\SI{120}{nm} with H$_3$PO$_4$ (5 {\% vol.}).

The growth conditions were similar to those described  by \citealt{bib-RUI2018}. For chemical modulations with a length below \SI{60}{nm}, the composition is Fe\(_{65}\)Ni\(_{35}\), while for lengths above \SI{60}{nm}, the composition is Fe\(_{80}\)Ni\(_{20}\). After the electrochemical growth, the AAO templates were dissolved in H\(_{3}\)PO\(_{4}\)(\SI{0.4}{M}) and H\(_{2}\)CrO\(_{4}\)(\SI{0.2}{M}). Then, the nanowires were dispersed on \SI{20}{nm}-thick  Si\(_{3}\)N\(_{4}\) windows to allow for transmission microscopy. To nucleate domain walls, nanowires were contacted electrically using laser lithography in order to allow the injection of \SI{}{\giga\hertz} bandwidth electric pulses. Resistivity values obtained are around \SI{20}{\micro\ohm\centi\metre}. To prevent sample oxidation and to increase thermal transfer to the Si\(_{3}\)N\(_{4}\), a \SI{20}{nm}-thick layer of Al$_2$O$_3$ was deposited on the device by Atomic Layer Deposition (ALD
). Additionally, \SI{100}{nm} of Al was deposited on the backside of the substrate.

The wires were imaged with X-ray Magnetic Circular Dichroism (XMCD) coupled to Transmission X-ray Microscopy (TXM) \cite{Sorrentino2015} at the MISTRAL beamline of ALBA synchrotron and by X-ray ptychography\cite{Pfeiffer2017,bib-DON2016,Mille2022} at the HERMES STXM beamline of SOLEIL synchrotron. The photon energy was set to the Fe L$_{3}$ absorption edge, and the sample holder could be rotated  from the plane normal to the X-ray incidence in order to be sensitive to both axial and azimuthal magnetization components. During TXM imaging, series of about 64 images per polarity were were acquired, registered and averaged in order to increase the signal-to-noise ratio, while keeping under control the loss of spatial resolution due to time drift. Ptychographic reconstruction was carried out using the open-source PyNX software developed at the European Synchrotron Radiation Facility \cite{FavreNicolin2020}. If not initially present, DWs were nucleated in situ by applying a \SI{10}{ns}-long pulse of current with amplitude \SI{1e12}{\ampere\per\meter\squared}. 

HAADF-STEM imaging and EDS mapping were performed using a double aberration-corrected Thermo Fisher Scientific Spectra 300 scanning transmission electron microscope (STEM) equipped with a Super-X EDS detector. The convergence semi-angle was set to 19.5 mrad. To improve the signal-to-noise ratio (S/N) of the EDS-STEM data, the spectrum imaging datasets were pre-filtered using a 3–5 pixel averaging filter prior to extracting the Ni and Fe net-count intensity maps. Composition profiles were subsequently obtained by integrating across nearly the entire nanowire diameter (integration widths ranging from 108 to 130 nm). The interface position was defined as the midpoint between the 10\% and 90\% intensity levels of the normalized Fe composition profile, while the interface width was defined as the distance between these two levels. To determine the width of the chemical defect interfaces, the Fe composition profiles were fitted using a Boltzmann sigmoid function. The interface width was thus defined as the distance over which the composition changes from 10\% to 90\% of the total compositional difference across the interface.

We performed micromagnetic simulations with the \textit{mumax3} code\cite{bib-VAN2014}. The system considered is a  cylindrical Permalloy nanowire of \SI{130}{nm} diameter and \SI{2}{\micro\meter}  length, with a \SI{60}{nm}-long or \SI{100}{nm}-long Fe\(_{80}\)Ni\(_{20}\) chemical modulation at its center. In order to mimic an infinite wire, the magnetic charges at the wire ends were removed numerically. The mesh size used was \SI{2}{nm}. The following material parameters were used\cite{bib-COU1996}: spontaneous magnetization $M_1 = \SI{8e5}{A/m}$ for the Permalloy segments, $M_2 = \SI{14e5}{{}A/m}$ for the chemical modulation, single exchange stiffness for all materials $A = \SI{1.3e-11}{J/m}$ and zero magnetocrystalline anisotropy. The choice of an identical value for exchange stiffness stems from the similarity of ordering temperature for the two compositions \cite{bib-COU1996}. The resulting dipolar exchange lengths are $\SI{5.6}{nm}$ and $\SI{3.3}{nm}$, respectively. To study static or quasi-static dynamics a damping term of $\alpha=1$ was used. The initial state considered was a BPW at z $=\SI{\pm500}{nm}$ and a chemical modulation centered at z $= 0$. Circulations of both the BPW and the modulation are defined arbitrarily with respect to $+\unitvectz$, and labeled $C+$ for positive and $C-$ for negative.

For the tracking of topological elements (\figref{fig:BPW-sw-under-field-topology}) characterizing the type of wall~(Bloch point for BPW, surface vortex/antivortex for the TVW), we use an in-house post-processing tool developed specifically for tetrahedron-based finite elements and applied here to a \SI{90}{nm} diameter wire for computational reasons \cite{AlvaroGomez2025topo}. These simulations are intended to elucidate the underlying physical mechanisms governing domain wall–modulation interactions, rather than to reproduce exact quantitative values.



\begin{acknowledgements}

This work received financial support from the French RENATECH network, implemented at the Upstream Technological Platform in Grenoble PTA  (ANR-22-PEEL-0015), and from the French National Research Agency (Grant No. ANR-17-CE24-0017 MATEMAC-3D; Grant ANR-22-CE24-0023 DIWINA), from the Spanish Research Agency (MCIN/AEI/10.13039/501100011033) through projects PID2024-155385NB-C31 and PID2024-155385NA-C32 and by Comunidad de Madrid through project TEC-2024/TEC-380 (Mag4TIC).  S. R.-G. acknowledge funding through Advanced Materials programme supported by MCIN with funding from the European Union NextGenerationEU (PRTR-C17.I1) and by the Generalitat de Catalunya.  We acknowledge support from the team of the Nanofab platform (CNRS N\'eel institut),  from the ALBA in-house research program and MISTRAL and CIRCE beamlines and from the SOLEIL HERMES beamline. Authors acknowledge the use of instrumentation as well as the technical advice provided by the Joint Electron Microscopy Center at ALBA (JEMCA) and funding from Grant IU16-014206 (METCAM-FIB) to ICN2 funded by the European Union through the European Regional Development Fund (ERDF), with the support of the Ministry of Research and Universities, Generalitat de Catalunya.

\end{acknowledgements}



\bibliography{references}

\end{document}


\title{Supplementary Information: Domain-wall circulation gating and pinning via compositional engineering of cylindrical nanowires}
\author{Laura \'Alvaro-G\'omez}
\email{laura.alvaro@ucm.es}
 \affiliation{Dpto. de F\'isica de Materiales, Universidad Complutense de Madrid, 28040 Madrid, Spain.}
 
\author{Claudia Fern\'andez-Gonz\'alez}
\affiliation{Alba Synchrotron Light Facility, CELLS, 08290 Cerdanyola del Vall\`es, Barcelona, Spain.}

\author{Sandra Ruiz-G\'omez}
\affiliation{Alba Synchrotron Light Facility, CELLS, 08290 Cerdanyola del Vall\`es, Barcelona, Spain.}

\author{Nicolas Mille}
\affiliation{Synchrotron SOLEIL, l'Orme des Merisiers, Saint-Aubin, FR-91192 Gif-sur-Yvette Cedex, France.}

\author{Jean Christophe Toussaint}
\affiliation{Univ. Grenoble Alpes, CNRS, Institut N\'eel, 38000 Grenoble, France.}

\author{Rachid Belkhou}
\affiliation{Synchrotron SOLEIL, l'Orme des Merisiers, Saint-Aubin, FR-91192 Gif-sur-Yvette Cedex, France.}

\author{Michael Foerster}
\author{Eva Pereiro}
\author{Luc\'ia Aballe}
\author{N\'uria Bagu\'es}
\affiliation{Alba Synchrotron Light Facility, CELLS, 08290 Cerdanyola del Vall\`es, Barcelona, Spain.}

\author{Christophe Thirion}
\affiliation{Univ. Grenoble Alpes, CNRS, Institut N\'eel, 38000 Grenoble, France.}

\author{Daria Gusakova}
\affiliation{Univ. Grenoble Alpes, CNRS, CEA,  SPINTEC, 38000 Grenoble, France.}

\author{Aurelien Masseboeuf}
\affiliation{Univ. Grenoble Alpes, CNRS, CEA, SPINTEC, 38000 Grenoble, France.}

\author{Olivier Fruchart}
\affiliation{Univ. Grenoble Alpes, CNRS, CEA,  SPINTEC, 38000 Grenoble, France.}

\author{Lucas P\'erez}
 \email{lucas.perez@ucm.es}
 \affiliation{Dpto. de F\'isica de Materiales, Universidad Complutense de Madrid, 28040 Madrid, Spain.}
 \affiliation{IMDEA Nanociencia, Campus de Cantoblanco, 28049 Madrid, Spain.}

\maketitle

\section{Detailed analysis of the compositional barriers}

\begin{figure}
\centering\includegraphics[width=0.8\columnwidth]{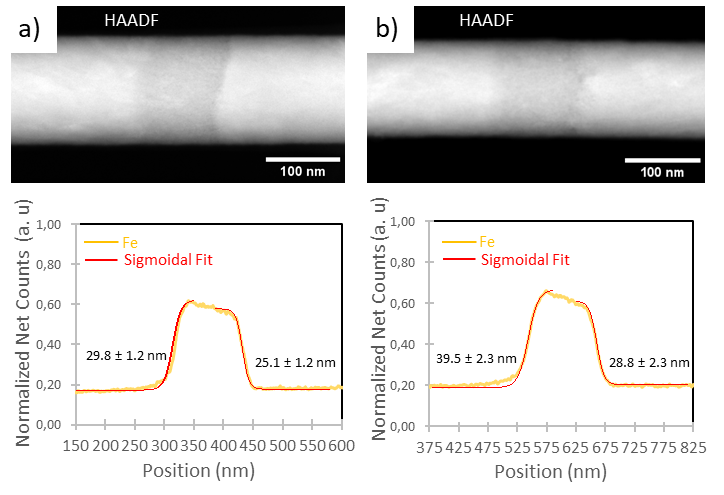}
\caption{HAADF-STEM images and Line profiles of normalized Fe net counts ratio together with the sigmoidal for the two type of modulations shown in the main manuscript.}

\label{fig:TEM2}
\end{figure}

Figure~\ref{fig:TEM2} presents HAADF-STEM images of two compositional modulations ($M1$ and $M2$) from different nanowires (NWs) together with the line profile of the normalized Fe net counts and the fitting obtained integrating across almost the full  nanowire diameter (integrated areas between 108 and 130 nm). These regions, which are representative of the chemical modulations observed throughout the sample, can be readily identified by distinct variations in the HAADF contrast. The corresponding EDS maps (Figure 2 of the main manuscript) reveal that the interfaces are not strictly perpendicular to the nanowire growth axis, but exhibit a curved or step-like morphology. The fine structural features cannot be fully resolved because the EDS signal is averaged over the entire nanowire thickness. The normalized Fe net-count profiles further show that the composition within each modulation presents a gradual decrease in Fe content from left to right of approximately 6\%. The interface position was defined as the midpoint between the 10\% and 90\% intensity levels of the normalized Fe composition profile, while the interface width was determined by fitting the profile with a Boltzmann sigmoid function and subsequently calculating the distance between the 10\% and 90\% composition levels. The resulting values reveal a clear asymmetry in the interface widths at the two boundaries of each modulation, reflecting the intrinsic asymmetry of the electrochemical growth process, namely the transition from high to low overpotential at one interface and from low to high overpotential at the other. Notably, both the interface widths and the local structural variations extend over only a few tens of nanometres, whereas the characteristic length scale of the magnetic textures is on the order of several hundreds of nanometres. Consequently, from the perspective of their interaction with the magnetic textures, the compositional modulations can be considered structurally homogeneous blocks. Finally, the total modulation widths, estimated from the midpoint positions of the two interfaces for $M1$ and $M2$, are $119 \pm 2\text{ nm}$ and $123 \pm 3\text{ nm}$, respectively.

\section{Micromagnetic simulation: scaling study}

To assess the robustness of the circulation reversal of a Bloch-point wall (BPW) upon propagation through a chemical modulation of antiparallel circulation, we performed a systematic scaling study using \textit{MuMax3}, considering nanowire diameters ranging from 80 nm to 150 nm and modulation lengths of 50 nm and 100 nm. The initial configuration consists of a BPW with a circulation antiparallel to that of the chemical modulation. An axial magnetic field is then applied in a stepwise manner using 2 ns-long field plateaus of increasing magnitude. In all simulated cases, the same sequence of events is observed: the domain wall approaches the modulation while the internal energy increases due to the repulsive interaction associated with the antiparallel circulation. The system eventually reaches a maximum energy configuration (the effective energy barrier), after which the BPW undergoes a circulation reversal and subsequently depins from the modulation once a critical field threshold is reached. These results demonstrate that, while the quantitative thresholds scale with the specific geometry, the qualitative behavior — namely circulation-selective pinning and deterministic switching — remains fully robust across the entire explored parameter range.

\begin{figure}
\centering\includegraphics[width=0.45\columnwidth]{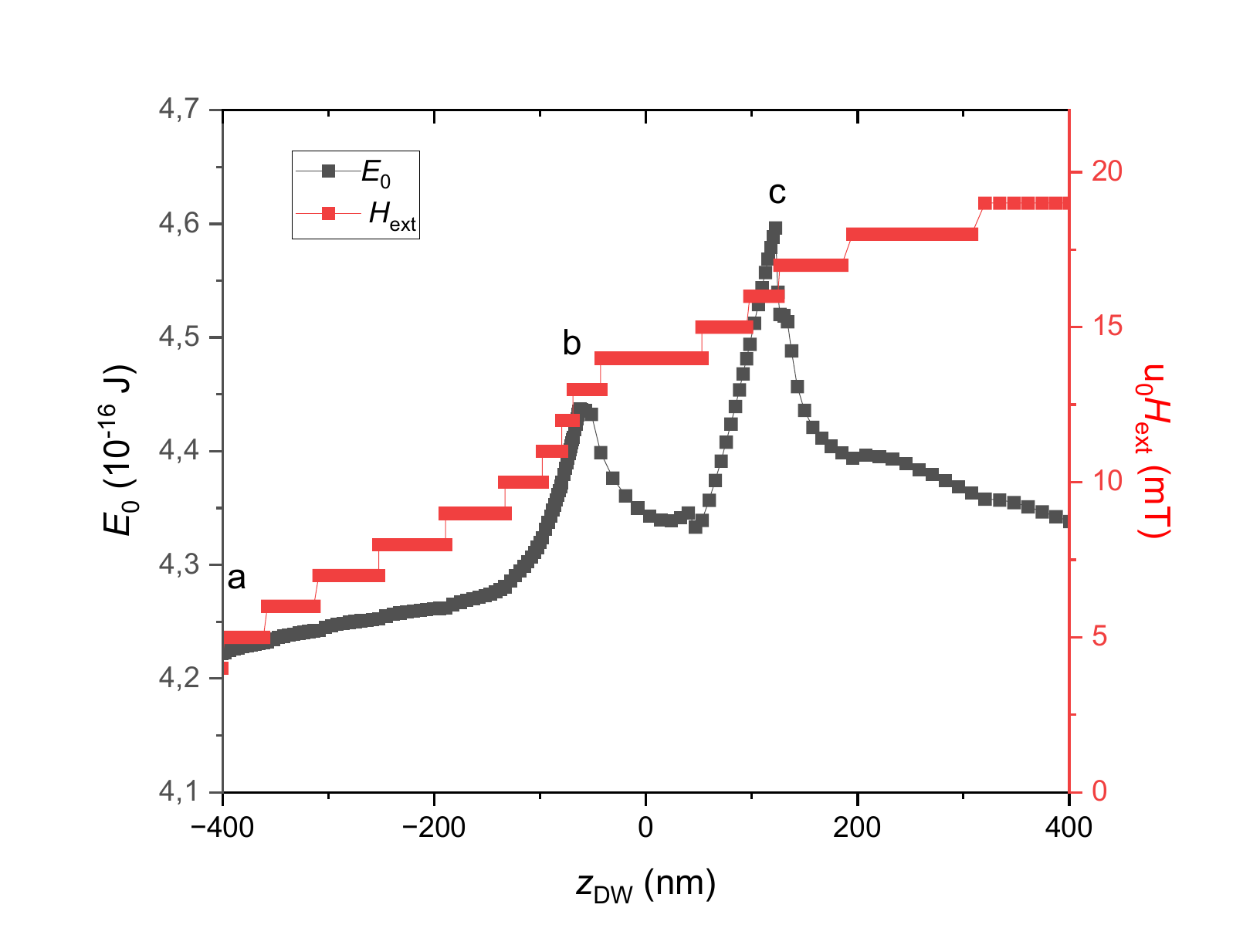}
\caption{Extraction of the relevant quantities from a representative micromagnetic simulation. 
The internal energy $E_0$ (black squares) and applied magnetic field $\mu_0 H_{\mathrm{ext}}$ (red symbols) 
are shown as a function of the domain wall position $z_{\mathrm{DW}}$.
The three quantities used in the scaling analysis are defined as follows:
(a) \textit{Initial energy}, corresponding to the energy of the domain wall far from the modulation;
(b) \textit{Barrier height}, defined as the maximum energy reached before the domain wall crosses the modulation;
(c) \textit{Depinning field}, defined as the value of the applied field at which the domain wall transitions from a pinned to a propagating regime.}

\label{fig:extract-quantities}
\end{figure}

To extract the relevant physical quantities (initial energy, barrier height, and depinning field), we analyze the evolution of both the internal energy and applied magnetic field during domain wall propagation. A representative example of this extraction procedure is depicted in Fig.~\ref{fig:extract-quantities}.  Here, the internal energy $E_0$ (black squares) and the applied magnetic field $\mu_0 H_{\mathrm{ext}}$  (red symbols) are plotted as a function of the domain wall position $z_{\mathrm{DW}}$ as the wall is driven towards and through the chemical modulation.

At large distances from the modulation (region a), the initial internal energy remains approximately constant, corresponding to the equilibrium state of the undisturbed domain wall. As the wall approaches the modulation, the energy gradually increases due to the repulsive interaction associated with the antiparallel circulation. This energetic rise continues until a peak is reached (region b), which defines the effective energy barrier for propagation through the modulation. Beyond this point, a further increase in the applied field drives a transition in the domain wall configuration, accompanied by a sharp decrease in internal energy as the wall propagates past the modulation. The depinning field (region c) is identified as the exact value of the applied field at which this structural transition occurs, marking the onset of irreversible domain wall motion beyond the pinned state.

\begin{figure}
\centering\includegraphics[width=0.5\columnwidth]{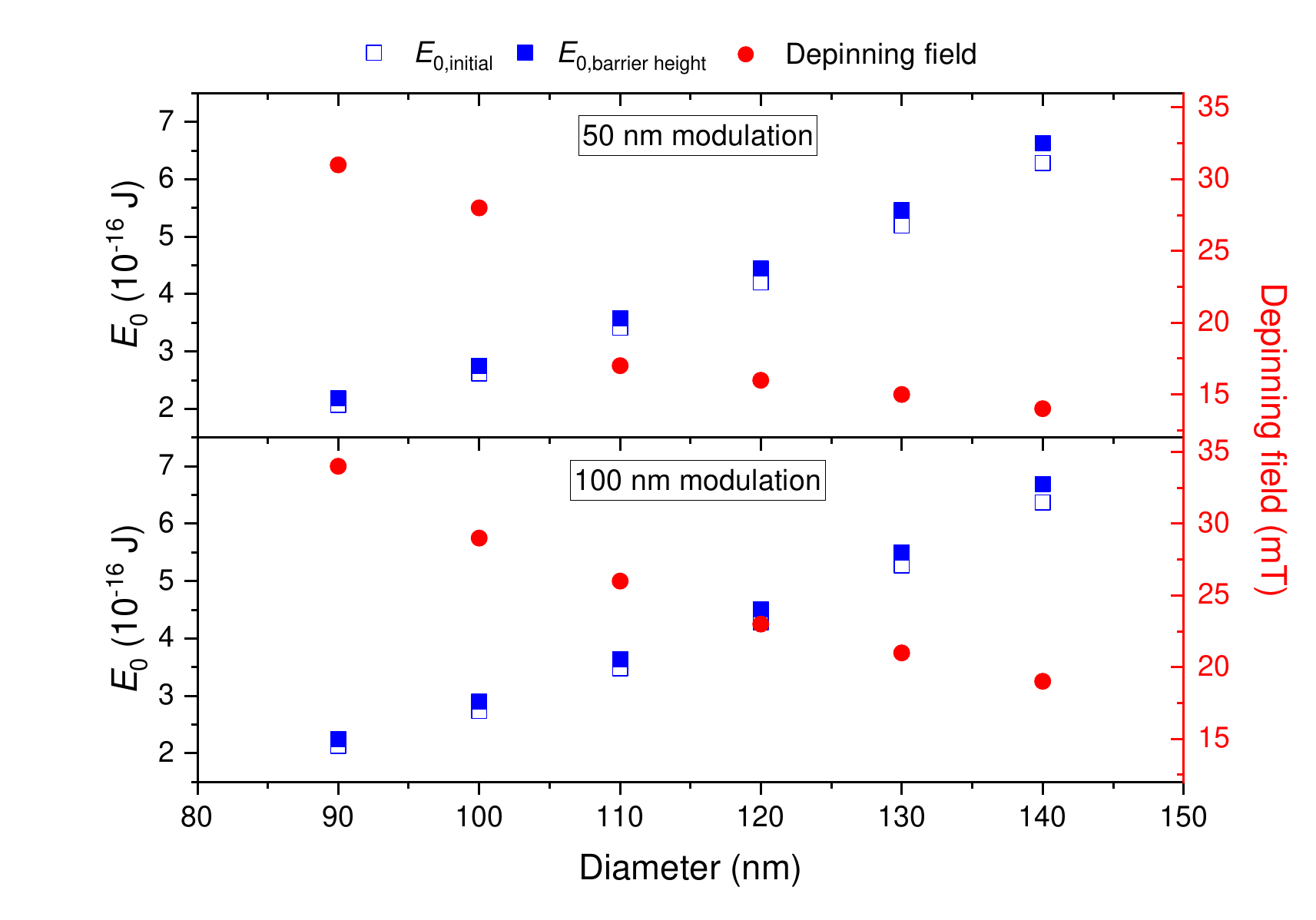}
\caption{Scaling study of the domain wall- modulation interaction as a function of nanowire diameter for modulation lengths of 50 nm (top panel) and 100 nm (bottom panel). 
The initial internal energy (open blue squares), barrier height (filled blue squares), and depinning field (red circles) are shown for each geometry.  }

\label{fig:quantitative-analysis}
\end{figure}

The resulting values of the initial internal energy (open blue squares), barrier height (filled blue squares), and critical depinning field (red circles) are summarized in Fig.~\ref{fig:quantitative-analysis}, which displays the evolution of these quantities as a function of the nanowire diameter for two distinct modulation lengths ($50\text{ nm}$ and $100\text{ nm}$). In both cases, the initial energy and the barrier height increase slightly with modulation length, reflecting the higher energetic cost associated with maintaining the curling configurations within longer segments. In contrast, the depinning field decreases with increasing diameter, indicating that systems with larger cross-sections require a weaker external field to overcome the energy barrier and drive the structural transition.

Importantly, despite these quantitative variations, the exact same qualitative behavior is consistently observed across all simulated geometries: the presence of a well-defined energy barrier associated with the antiparallel circulation, followed by a deterministic circulation switching event and subsequent domain wall depinning. This comprehensive scaling analysis confirms that the underlying physical mechanism is remarkably robust and remains invariant over the entire experimentally relevant parameter range.

\bibliography{references}